\documentclass[aps,superscriptaddress,showpacs,nofootinbib,floatfix]{revtex4}
\usepackage{amsfonts,amscd,amsmath,amssymb,graphicx,color}
\def\bv{T_{\text{v}}}
\newcommand{\oh}{\frac{1}{2}}
\def\ep{\text{e}}
\begin{document}
\preprint{LMU-ASC 109/10}
\title{Some Comments on Relativistic Hydrodynamics and Fuzzy Bags}
\author{Oleg Andreev}
\affiliation{Arnold Sommerfeld Center for Theoretical Physics, LMU-M\"unchen, Theresienstrasse 37, 80333 M\"unchen, Germany}
\affiliation{L.D. Landau Institute for Theoretical Physics, Kosygina 2, 119334 Moscow, Russia}
\begin{abstract} 
Three subjects are considered here: the relativistic hydrodynamics equations for a boost-invariant expanding fluid; the fuzzy bag model for the pressure which recently appeared in QCD phenomenology; and the early space-time evolution of the QCD matter, drawn from model studies, which can also be 
expected to arise in realistic fluid dynamics relevant to heavy ion collisions at LHC.
\end{abstract}
\pacs{25.75-q, 12.38.Mh, 12.38.Lg}
\maketitle


\section{Introduction}

The idea of exploiting the laws of hydrodynamics to describe the expansion of the strongly interacting matter created in high energy collisions was pioneered by Landau in 1953 \cite{dau}. In recent years, data from the Relativistic Heavy Ion Collider (RHIC) provided striking 
evidence for a strong collective expansion that is in good quantitative agreement with hydrodynamic predictions.\footnote {For reviews, see \cite{rev}.} Moreover, these data emphasized the exciting possibility that the QCD matter 
created in the collisions shows properties of a near-perfect fluid. Once well-calibrated ideal fluid dynamical benchmarks have been established under appropriate experimental conditions, deviations from perfect fluid behavior can be used to explore transport properties, such as viscosities and heat conductivities of the QCD matter. If unambiguously extracted from data, they are prime candidates for the next compilation of the Particle Data Group 
\cite{wid}. Such efforts are among those that define the present forefront of research in heavy-ion collision dynamics. 

The results at RHIC also indicate that the matter created in heavy ion collisions behaves like a strongly coupled liquid rather than a weakly interacting plasma of quarks and gluons. In this case, the MIT bag model that effectively includes strong interactions via a bag constant can be of use as an equation of state. Recently, the bag model has taken an interesting turn with the detailed analysis \cite{pis} of the lattice data \cite{lattice-K}. The surprise of this analysis is that at zero chemical potential it reveals a term quadratic in temperature as the leading correction to the ideal gas term in the pressure.\footnote{Phenomenologically, such a term was first suggested in \cite{T2}.} The recent data \cite{lattice,pan,baz} also indicate that this is generic: that the equation of state of the original bag model gets modified as \footnote{Rob Pisarski called it a ‘‘fuzzy’’ bag model for the pressure \cite{pis}.}

\begin{equation}\label{eofs}
	p(T)=a(T^4-\alpha T^2)-B
	\,,\quad\text{with}\quad
	T_{\text{\tiny min}}<T<T_{\text{\tiny max}}\,.
\end{equation}

As noted, the novelty is the $T^2$ term, while the remaining two are the standard bag terms, with B a bag constant and $a$ a parameter. A common choice is to take $a$ from perturbation theory up to one loop order. $T_{\text{\tiny min}}$ is close to a critical temperature $T_c$ (or some approximate ”$T_c$” for a crossover). A small difference between $T_c$ and $T_{\text{\tiny min}}$ may vary with the model. $T_{\text{\tiny max}}$ is set by perturbation theory such that to leading orders it is applicable only for temperatures higher than $T_{\text{\tiny max}}$. In fact, there are arguments in the literature \cite{q2-rev} that in QCD long perturbative series (or the UV renormalons) result in the so-called quadratic corrections. From this point of view, the $T^2$ term in \eqref{eofs} is nothing else but an example of the quadratic correction. In the dual (string) description, these 
quadratic corrections are parameterized in terms of the metric in extra dimensions and do result in the $T^2$ term for the pressure \cite{ads}. 

What is the physical significance of such a modification? In the present paper, we look into how it affects the early space-time evolution of the QCD matter and why it may be relevant to some attempts to extract dissipative transport coefficients from data. 

In general, the $3+1$-dimensional hydrodynamic equations are very complicated and require powerful numerical simulations. Here, we consider a simplified model which, however, is a good approximation to the early time evolution of the QCD matter. It incorporates several characteristic features of heavy ion collisions, while lacking many of the technical complications of the most general case. Moreover, it allows us to find analytic solutions. This is the fluid dynamical model proposed by Bjorken \cite{bj}.

In this $1+1$-dimensional model, introducing the Milne variables proper time $\tau=\sqrt{t^2-z^2}$ and space-time rapidity $\xi=\text{arctanh}(z/t)$, boost invariance simply translates to requiring all hydrodynamic variables to be independent of rapidity. As a result, the hydrodynamic equations 
become exceptionally simple \cite{rev}. To first order in the gradient expansion, one has 

\begin{equation}\label{hydro}
\tau\frac{d\varepsilon}{d\tau}=-\varepsilon -p+\Phi-\Pi
	\,
\end{equation}
with 
\begin{equation}\label{scalars}
	\Phi=\frac{4\eta}{3\tau}\,,\qquad
	\Pi=-\frac{\zeta}{\tau}\,
\end{equation}
given by their Navier-Stokes values. Here $\eta$ and $\zeta$ are the shear and bulk viscosity coefficients, respectively.

In order for the equations \eqref{hydro} and \eqref{scalars} to be closed, one has to specify an equation of state  $p(\varepsilon)$ and expressions relating $\eta$ and $\zeta$ to $\varepsilon$. The simplest case to be considered is that of a perfect fluid, $\eta=\zeta=0$, with a bag model type equation of state, eq.\eqref{eofs} at $\alpha=0$. If so, then the evolution equation \eqref{hydro} can be solved analytically. One gets 

\begin{equation}\label{Bjorken}
	T(\tau)=T_0\Bigl(\frac{\tau_0}{\tau}\Bigr)^{\frac{1}{3}}
	\,,
\end{equation}
with the initial condition $T(\tau_0)=T_0$. This is the celebrated Bjorken solution \cite{bj} which defines the baseline on top of which dissipative and fuzzy bag effects have to be established. Certainly, deviations must be sufficiently small for the gradient expansion underlying dissipative hydrodynamics to be valid. 

\section{Perfect Fluid}

In the absence of dissipative effects, the entropy per unit of rapidity is a constant of the motion \cite{bj}. Therefore the entropy density is 

\begin{equation}\label{s}
	s(\tau)=s_0\frac{\tau_0}{\tau}
	\,.
\end{equation}

For the fuzzy bag \eqref{eofs}, we can still obtain an explicit formula for $T(\tau)$ by solving a cubic equation. So we get

\begin{equation}\label{T}
T(\tau)=
\begin{cases}
\bigl(\frac{c}{\tau}\bigr)^{\frac{1}{3}}
\Bigl[\Bigl(\oh+\oh\sqrt{1-\frac{\alpha^3}{54}\bigl(\frac{\tau}{c}\bigr)^2}\Bigr)^{\frac{1}{3}} 
+\Bigl(\oh-\oh\sqrt{1-\frac{\alpha^3}{54}\bigl(\frac{\tau}{c}\bigr)^2}\Bigr)^{\frac{1}{3}}\Bigr]\quad&
\text{if $\tau\leq\sqrt{\frac{54}{\alpha^3}}c$}\,,\\
\sqrt{\frac{2}{3}\alpha}\,
\cos\Bigl[\frac{1}{3}\arccos\Bigl(\sqrt{\frac{54}{\alpha^3}}\frac{c}{\tau}\Bigr)\Bigr]\quad&\text{if $\tau\geq\sqrt{\frac{54}{\alpha^3}}c$ }\,,
\end{cases}
\end{equation}
where $c=\tau_0T_0(T^2_0-\oh\alpha)$. The solution \eqref{Bjorken} emerges in the limit $\alpha\rightarrow 0$, as expected.

In Figure 1 we show the proper time evolution of the temperature and energy density for the bag and fuzzy bag models. Following our discussion on the fuzzy bag, we choose the initial temperature to correspond to $T_{\text{\tiny max}}$ and stop the evolution at $T=T_{\text{\tiny min}}$. It is expected on general grounds that $T_{\text{\tiny min}}$ is somewhere between $0.2\,\text{GeV}$ and $0.3\,\text{GeV}$, while $T_{\text{\tiny max}}$ is around $1\,\text{GeV}$. This is consistent with the recent lattice data for QCD with almost physical quark masses in the interval $0.3\,\text{GeV}\leq T\leq 0.8\,\text{GeV}$ \cite{lattice}. So, we take the initial temperature $T_0$ to be $0.8\,\text{GeV}$. It is worth noting that this value is expected to be reached in the coming years as the LHC goes into its heavy ion program. Given $T_0$, the initial time can be estimated by using the uncertainty principle \cite{kapustaMc}: $\tau_0\langle E\rangle_0\sim 1$, where $\langle E\rangle_0\sim 3 T_0$. This results in $\tau_0=0.08\,\text{fm/c}$. Thus, we consider the early time evolution where the Bjorken model is a good approximation. The value of $\alpha$ is fixed from the slopes for the Regge trajectory of $\rho (n)$ mesons and the linear term of the Cornell potential \cite{A-q2}. For $N_c=N_f=3$, it is given by $\alpha\approx 0.09\,\text{GeV}^2$. Finally, the value of the bag constant is set to $(0.2\,\text{GeV})^4$.  

\begin{figure}[ht]
\centering
\includegraphics[width=7.1cm]{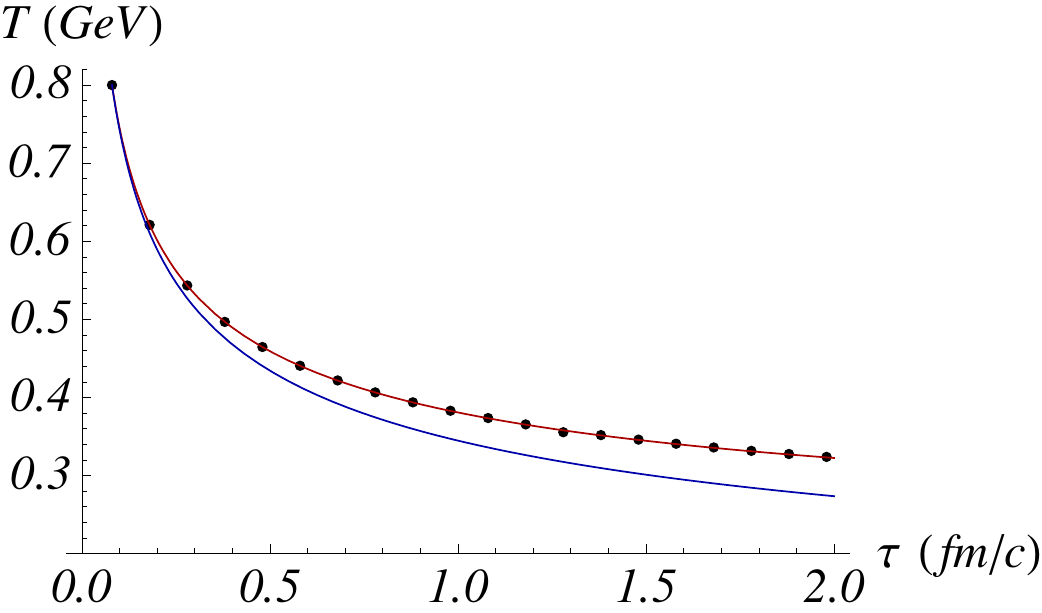}
\hfill
\includegraphics[width=6.9cm]{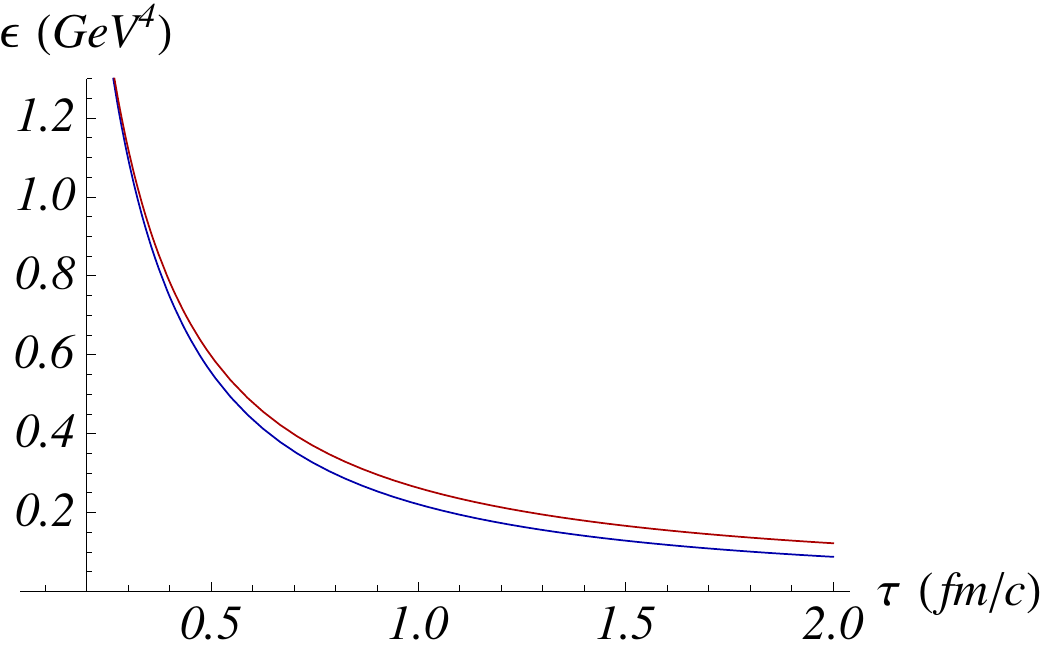}
\caption{\small{Left: Evolution of the temperature. Right: Evolution of the energy density. In both the cases, the lower blue and upper red curves correspond to the MIT and fuzzy bag models, respectively. The dots come from the approximation \eqref{Tshort}.}} 
\end{figure}
What we see from this Figure is that the temperature decreases with the proper time more slowly in the presence of the $T^2$ term in the equation of 
state. Note that the bag constant has no effect on the evolution of the temperature. Thus, the Bjorken solution is also valid for a conformal fluid. The 
maximum discrepancy occurring at $\tau=2\,\text{fm/c}$ is of order $13\%$. It rapidly decreases with decreasing $\tau$ and becomes almost negligible for $\tau<0.3\,\text{fm/c}$. 

For practical purposes, the expression \eqref{T} looks somewhat awkward. A possible way out is to expand it in a series and then see where that takes us. If we ignore all higher terms in the small $\alpha$ expansion, which is equivalent to the small $\tau$ expansion, then a final result can be 
written in a simple form 

\begin{equation}\label{Tshort}
	T(\tau)\approx\Bigl(\frac{c}{\tau}\Bigr)^{\frac{1}{3}}+\frac{\alpha}{6}\Bigl(\frac{\tau}{c}\Bigr)^{\frac{1}{3}}
	\,.
\end{equation}
Here the second term represents the leading $\alpha$-correction to the Bjorken solution. Since its value is positive, the slowing down of the 
temperature decrease occurs (at least for small $\tau$). In Figure 1 we have plotted the proper time dependence of the temperature as it follows from \eqref{Tshort}. Remarkably, with our initial conditions, the difference between \eqref{T} and \eqref{Tshort} is negligible in the interval $0.08\,\text{fm/c}\leq\tau\leq 2\,\text{fm/c}$.

Finally, let us discuss the evolution of the energy density. In the fuzzy bag model it is given by $\epsilon=3aT^4-a\alpha T^2+B$.\footnote{For completeness, we have included a formula for $\epsilon(\tau)$ in the Appendix.} Because of the minus sign in front of the $T^2$ term, one would expect that the energy density will fall faster in the presence of the $T^2$ term. In reality, the effect of the slowing down of the evolution of $T$ dominates and, as a consequence, the energy density falls slower in the fuzzy bag model than in the original one. We illustrate this in the right panel of 
Figure 1.

\section{Viscous Fluid}

One of the most fascinating developments in recent years has been the conjecture \cite{kss} that there may be a fundamental bound from below 
on the ratio of the shear viscosity $\eta$ to the entropy density $s$ \footnote{We use units where $\hbar=c=k_{\text{\tiny B}}=1$.}

\begin{equation}\label{kss}
	\frac{1}{4\pi}\leq\frac{\eta}{s}
	\,.
\end{equation} 
Given the conjectured existence of the bound, it is natural to ask how closely does the QCD matter (fluid) produced at RHIC (LHC) approach the 
bound. While some methods provide consistent support for a conclusion that the produced matter is within a factor of $2-3$ above the conjectured bound \cite{zajc}, controversy still remains \cite{roma}. We now wish to understand whether the modification of the original bag model will affect the conclusion.

\subsection{Shear Viscosity or Modified Equation of State?}

In the case of the MIT bag model, the solution of the evolution equation \eqref{hydro} is known analytically for $\zeta=0$ and $\eta/s=const$ \cite{Bv}

\begin{equation}\label{conk}
	T(\tau)=T_0\Bigl(\frac{\tau_0}{\tau}\Bigr)^{\frac{1}{3}}+\frac{2\kappa}{3\tau_0}
	\biggl[\Bigl(\frac{\tau_0}{\tau}\Bigr)^{\frac{1}{3}}-\frac{\tau_0}{\tau}
	\biggr]
	\,,
\end{equation}
where $\kappa=\eta/s$. The second term is a viscous correction. It shows that the temperature decreases with the proper time more slowly in the presence of the shear viscosity. 

Now, a problem arises. For $\tau<\tau_0/(\frac{1}{3}+\frac{\tau_0}{2\kappa}T_0)^{3/2}$, the function \eqref{conk} grows with increasing $\tau$. This is unphysical. What it means is that dissipative effects are large and, as a consequence, the first order theory is not applicable any more.\footnote{This 
is an example of reheating. For more discussion, see \cite{wid} and references therein.} To proceed, therefore, we need to pick the initial time 
in such a way that $\tau_0\geq\tau_0/(\frac{1}{3}+\frac{\tau_0}{2\kappa}T_0)^{3/2}$. Together with $\tau_0 T_0=1/3$ \cite{kapustaMc}, this results in the 
following constraint $\kappa\leq 1/4$. Since the value of $1/(4\pi)$ is approximately $0.08$, it obeys the constraint. Moreover, it is possible to consider larger values of $\kappa$ (up to a factor of $3$).  Finally, let us note that the situation is opposite for the solution \eqref{T}, where 
the problem of small $\tau$ doesn't occur.

From our discussion so far, when the Bjorken solution \eqref{Bjorken} defines a baseline, the slowing down of the evolution of the temperature occurs if either the equation of state gets modified or the shear viscosity coefficient is not set to zero. A natural question arises: what is dominant? 

We can gain some understanding of this by plotting the curves given by equations \eqref{T} and \eqref{conk}. The results are shown in Figure 2. 
\begin{figure}[ht]
\centering
\includegraphics[width=8.3cm]{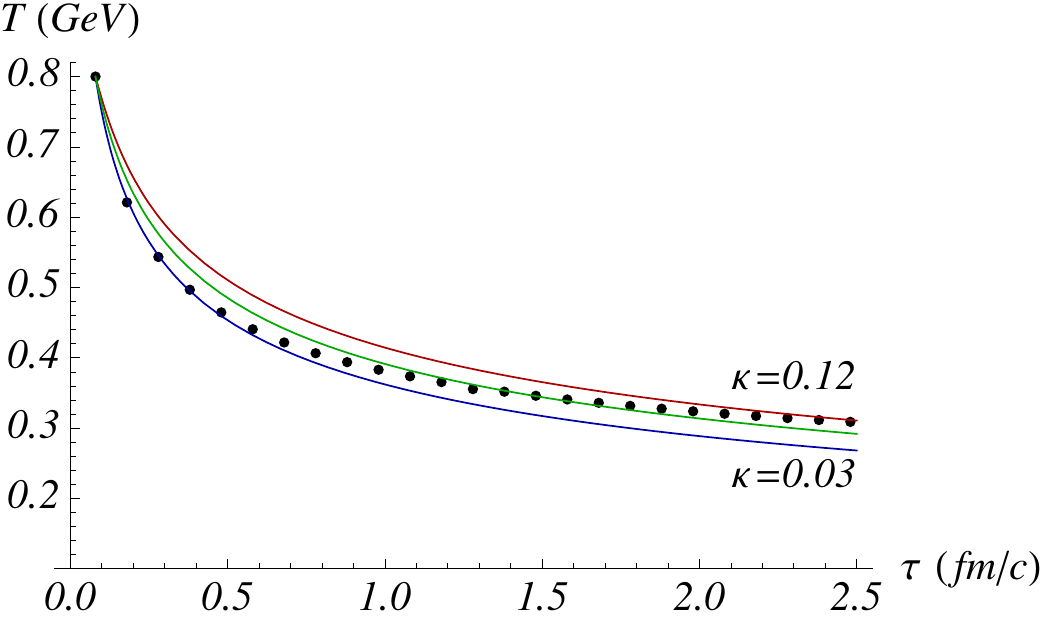}
\caption{\small{Evolution of the temperature. The green curve corresponding to $\kappa=1/(4\pi)$ lies between those for $\kappa=0.03$ and 
$\kappa=0.12$. The dots represent the solution \eqref{T} with $\alpha=0.09\,\text{GeV}^2$. As before, the initial conditions are $T_0=0.8\,\text{GeV}$ and $\tau_0=0.08\,\text{fm/c}$.}} 
\end{figure}
We see that the function \eqref{T} decreases with $\tau$ more slowly. Indeed, at the beginning of the evolution process it is comparable with the function \eqref{conk} at values of $\kappa$ below $1/(4\pi)$, while at the end it becomes comparable with that at greater values exceeding $1/(4\pi)$.
Also note that for $\kappa=1/(4\pi)$ the functions are almost indistinguishable in the interval $1.2\,\text{fm/c}\lesssim\tau\lesssim 1.7\,\text{fm/c}$. 

In fact, it is easy to understand why it is so. The reason for this is contained in the formulas \eqref{T} and \eqref{conk}. The function \eqref{T} goes to $\sqrt{\alpha/2}$ as $\tau\rightarrow\infty$, while the function \eqref{conk} goes to $0$. We will return to this issue in section VI.   

Thus, what we have learned is that these effects have the same sign and both are significant in the interval of primary interest 
$0.08\lesssim\eta/s\lesssim 0.24$. We consider the early time evolution $0.08\,\text{fm/c}\leq\tau\leq 2-2.5\,\text{fm/c}$, where the Bjorken model is 
a good approximation. In such a situation, our conclusion is likely to be reliable in heavy ion collisions as well. If so, it should be taken seriously by those who will try to extract the value of the shear viscosity from the LHC data and shed some light on the conjectured bound.

\subsection{Attempt of Synthesis}

The formula \eqref{T} is oversimplified for various reasons. For one thing, the shear and bulk viscosity coefficients must be included in the evolution equation. A common way to do so is to set $\eta/s$ to a constant and $\zeta$ to zero \cite{rev}. For the window 
$0.3\,\text{GeV}\leq T\leq 0.8\,\text{GeV}$, it gives a good approximation for the evolution. But this level of accuracy is sufficient nether for the 
next compilation of the Particle Data Group nor for shedding the light on the lower bound \eqref{kss}.

First, we will discuss the shear viscosity. Like in atomic and molecular physics, the ratio $\eta/s$ depends on $T$. In the temperature interval of interest it is an increasing function. Although there is 
strong evidence for this from lattice simulations \cite{lattice-shear}, the limited data don't allow one to suggest any reasonable parametrization. On the other hand, in an alternative approach based on the soft wall metric model of AdS/QCD a simple formula was derived in \cite{kapusta-shear}. For $T>T_c$, it is 

\begin{equation}\label{svis}
	\frac{\eta}{s}=\kappa\Bigl(1-k\frac{\alpha}{T^2}\Bigr)
	\,,
\end{equation}
where $k$ is a positive constant and $\alpha$ is given by \eqref{eofs}. The reason for this is that the model was also used to derive the equation of 
state \cite{ads}. In contrast to \cite{kapusta-shear}, we have not normalized \eqref{svis} to $\kappa=1/(4\pi)$.

The situation seems to be better with the bulk viscosity. According to \cite{DimaK}, $\zeta$ can be related with the QCD trace anomaly, yielding for high temperatures $\zeta/s\sim 1/T^2$. This conclusion is attractive because it agrees with what can be drawn from lattice simulations \cite{lattice-bulk}. 
In fact, the lattice data are well fitted by \cite{raj}

\begin{equation}\label{bvis}
	\frac{\zeta}{s}=m\frac{\alpha}{T^2}
	\,.
\end{equation}
Here $m$ is a positive constant and $\alpha$ is introduced for dimensional reasons.

This state of affairs, together with the evolution equation \eqref{hydro}, means that we can combine the viscous terms as
 
\begin{equation}\label{f}
	\upsilon=\frac{\eta}{s}+\frac{3\zeta}{4s}=\kappa\Bigl(1-\delta\frac{\alpha}{T^2}\Bigr)
	\,,
\end{equation}
where $\delta=k-3m/(4\kappa)$. 

There is an obvious question. Can the ratio $\zeta/s$, which is dependent of $T$, cancel the $T$ dependence of $\eta/s$? Note that this is not merely a 
problem with the choice of fitting functions. It is a more general phenomenological problem. The ratio $\eta/s$ is an increasing function of $T$, 
while the ratio $\zeta/s$ is a decreasing function of $T$. Therefore, if one combines these two, a cancellation among "$T$ dependencies" can happen. 

We have no real resolution of this problem. All what we can do is to illustrate it with an example. We do so in Figure 3. Here we take $k=1/4$ at $\kappa=1/(4\pi)$ as it follows from the results of \cite{kapusta-shear} and \cite{ads}. It is worth noting that $\eta/s\approx 0.06$ at $T=0.3\,\text{GeV}$ that is of course below the conjectured lower bound on $\eta/s$ of \cite{kss} but above $0.03$ of \cite{roma}. With the same values of parameters in \eqref{bvis} as those of \cite{raj}, we have $m\approx 0.024$ at $\alpha=0.09\,\text{GeV}$. Then we make an estimate of $\delta\approx 0.024$. Surprisingly, it is one order of magnitude smaller than $k$. We see therefore that $\upsilon$ is a slowly varying function of $T$:
\begin{figure}[ht]
\centering
\includegraphics[width=7.1cm]{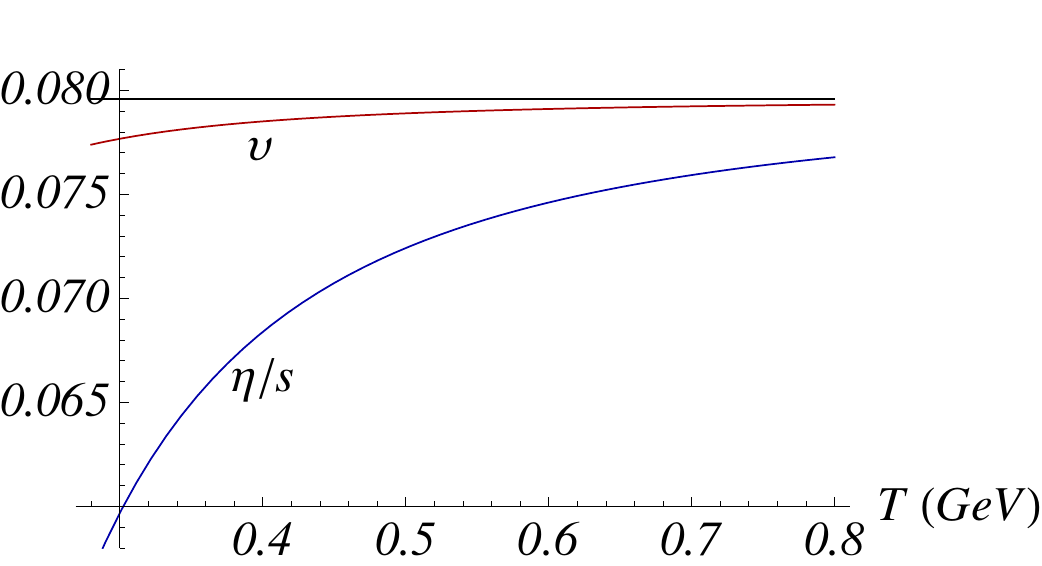}
\caption{\small{Values of $\upsilon$ and $\eta/s$ versus $T$. The horizontal line represents the conjectured lower bound on 
$\eta/s$.}} 
\end{figure}
there is a small difference between the conjectured value of $\kappa=1/(4\pi)$ and $\upsilon$. The maximum discrepancy occurring at $T=0.3\,\text{GeV}$ is of order $2\%$. For comparison, note that it becomes of order $25\%$ for $\eta/s$.

Having determined the viscosity coefficients, we can now consider the evolution equation. Combining \eqref{eofs} and \eqref{f}, 
we get 

\begin{equation}\label{hydro1}
6T'+\frac{2}{\tau}T-\Bigl(T'+\frac{1}{\tau}T\Bigr)\frac{\alpha}{T^2}-
\frac{4\kappa}{3\tau^2}\Bigl(2-\frac{\alpha}{T^2}\Bigr)
\Bigl(1-\delta\frac{\alpha}{T^2}\Bigr)
=0
	\,.
\end{equation}

We cannot unfortunately solve this equation analytically but we can do it numerically. Figure 4 shows 
\begin{figure}[ht]
\centering
\includegraphics[width=7cm]{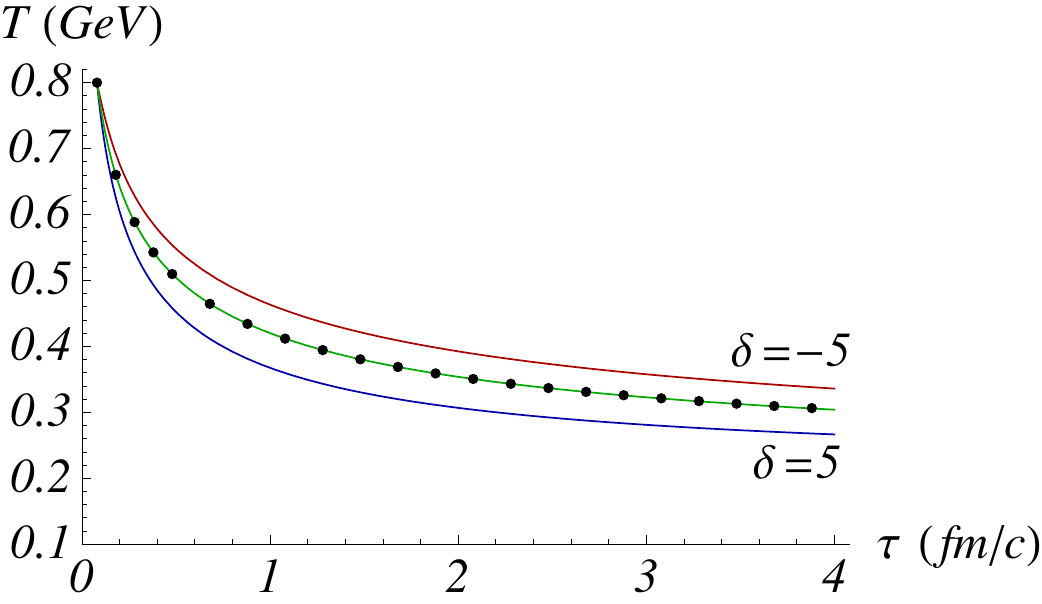}
\hfill
\includegraphics[width=7cm]{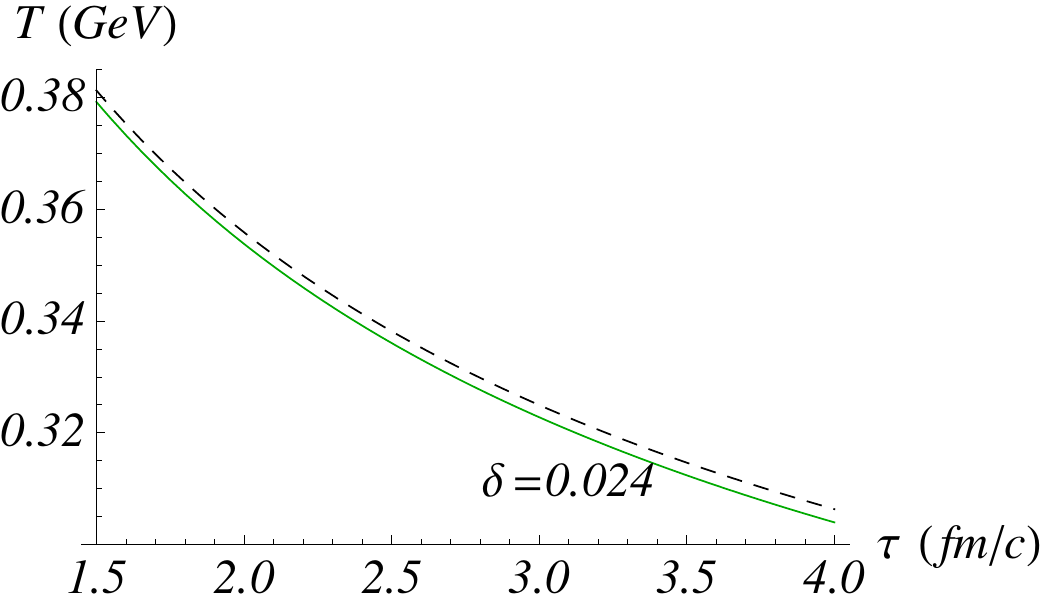}
\caption{\small{Evolution of the temperature. Left: The curve corresponding to $\delta=0.024$ lies between those 
for $\delta=\pm 5$. The dots come from the approximation \eqref{T1} at $\delta=0.024$. Right: The solid curve corresponds to $\delta=0.024$. 
The dashed curve represents the approximation \eqref{simple}.}} 
\end{figure}
the time dependence of the temperature for our initial conditions $T_0=0.8\,\text{GeV}$, $\tau_0=0.08\,\text{fm/c}$. We also take $\alpha=0.09\,\text{GeV}^2$ and $\kappa=1/4\pi$. In fact, to see significant effects on $T(\tau)$, the absolute value of the parameter $\delta$ must be increased by two orders of magnitude. This suggests that we can look for a solution which is a double power series in $\alpha$ and $\kappa$. 

We begin by expanding the solution in powers of $\alpha$. To first order, we get

\begin{equation}\label{T1}
	T(\tau)\approx\bv+\frac{\alpha}{6}\biggl[
	\frac{1}{\bv}-\frac{2\kappa}{3c}(1+6\delta)\Bigl(\frac{c}{\tau}\Bigr)^{\frac{1}{3}}
	\Bigl(\ln(\tau\bv)
-\frac{2\kappa}{3\tau\bv}\Bigr)\biggr]
\,
\end{equation}
with 
\begin{equation}
	\bv (\tau)=\Bigl(\frac{c}{\tau}\Bigr)^{\frac{1}{3}}-\frac{2\kappa}{3\tau}
	\,.
\end{equation}
As seen from the left panel of Figure 4, for $\kappa=1/(4\pi)$ it is a good approximation in the interval $0.08\,\text{fm/c}\leq\tau\leq 4\,\text{fm/c}$. 

Next, we expand \eqref{T1} in powers of $\kappa$. In this case, neglecting higher order terms gives   

\begin{equation}\label{simple}
	T(\tau)\approx\Bigl(\frac{c}{\tau}\Bigr)^{\frac{1}{3}}-\frac{2\kappa}{3\tau}+\frac{\alpha}{6}\Bigl(\frac{\tau}{c}\Bigr)^{\frac{1}{3}}
	\,.
\end{equation}
Again, as seen from the right panel of Figure 4, for $\kappa=1/4\pi$ it is a good approximation whose error is less then $0.9\%$ for the temperature evolution.

\section{Other Analytical Solutions}

Now we want to show that, with different choices of $\upsilon$, more solutions can be found analytically. These seems to be mainly of 
academic interest. 

Let us first consider the MIT bag model with 

\begin{equation}\label{vi}
	\upsilon=\frac{3\beta}{4T^2}
	\,.
\end{equation}
Here $\beta$ is a positive constant. One can think of this as a consequence of the fact that the model has $\eta=0$ and $\zeta\sim T$. The solution 
to the evolution equation is easily found to be 

\begin{equation}\label{T-kapusta}
	T(\tau)=T_0\Bigl(\frac{\tau_0}{\tau}\Bigr)^{\frac{1}{3}}
	\biggl(1+\frac{\beta}{\tau_0T^3_0}\ln\frac{\tau}{\tau_0}\biggr)^{\frac{1}{3}}
	\,.
\end{equation}
In fact, it can be added to a list of those in \cite{kapusta-rev} as a special case $n=-1$. Note also that \eqref{T-kapusta} decreases with $\tau$ more slowly than \eqref{Bjorken}.

A simple analysis shows that for $\tau<\tau_0\ep^{1-\tau_0T^3_0/\beta}$ the function \eqref{T-kapusta} grows with increasing $\tau$. It gives another 
example of reheating. A way to avoid this problem is to pick $\tau_0$ so that $\tau_0\geq\tau_0\ep^{1-\tau_0T^3_0/\beta}$. Together with 
$\tau_0T_0=1/3$ \cite{kapustaMc}, this results in the following constraint $T_0^2\geq 3\beta$.

Now let us explore what happens when the fuzzy bag model is accompanied by \footnote{Curiously, a linear dependence of $\eta/s$ on $T$ was 
derived in \cite{kapusta-shear} within the hard wall model of AdS/QCD.}

\begin{equation}\label{v-a}
	v=\frac{3}{4}\gamma T
	\,,
\end{equation}
where $\gamma$ is a positive constant. In this case, the evolution equation is integrated to give the entropy density as a function of the 
proper time 

\begin{equation}\label{s-A}
	s(\tau)=4a\frac{\hat c}{\tau}\ep^{-\frac{\gamma}{\tau}}
	\,,
\end{equation}
where $\hat c=\tau_0T_0(T^2_0-\oh\alpha)\ep^{\frac{\gamma}{\tau_0}}$. As before, we can obtain an explicit formula for $T(\tau)$ by solving a cubic 
equation. It is now 

\begin{equation}\label{T-A}
	T(\tau)=
	\begin{cases}
	\bigl(\frac{\hat c}{\tau}\bigr)^{\frac{1}{3}}\ep^{-\frac{\gamma}{3\tau}}
	\Bigl[\Bigl(\oh+\oh\sqrt{1-\frac{\alpha^3}{54}\bigl(\frac{\tau}{\hat c}\bigr)^2\ep^{\frac{2\gamma}{\tau}}}\Bigr)^{\frac{1}{3}} 
	+\Bigl(\oh-\oh\sqrt{1-\frac{\alpha^3}{54}\bigl(\frac{\tau}{\hat c}\bigr)^2\ep^{\frac{2\gamma}{\tau}}
	}\Bigr)^{\frac{1}{3}}\Bigr]
	& \text{if $\tau\ep^{\frac{\gamma}{\tau}}\leq\sqrt{\frac{54}{\alpha^3}}\hat c$}\,,\\
	\sqrt{\frac{2}{3}\alpha}\,
	\cos\Bigl[\frac{1}{3}\arccos\Bigl(\sqrt{\frac{54}{\alpha^3}}\frac{\hat c}{\tau}\ep^{-\frac{\gamma}{\tau}}\Bigr)\Bigr] 
	&\text{if $\tau\ep^{\frac{\gamma}{\tau}}\geq\sqrt{\frac{54}{\alpha^3}}\hat c$ }\,.
	\end{cases}
	\end{equation}

Some comments about formula \eqref{T-A} are in order. Let us first notice that for sufficiently large $\tau$ this solution approaches that of perfect fluid, with $c$ replaced by $\hat c$. Thus, viscous effects become negligible on long time scales. Next, there is apparently a problem of small 
$\tau$. In this case, one must take $\tau_0\geq\gamma$ in order to avoid any unphysical reheating. Finally, if we set $\alpha=0$, that means a bag model 
type equation of state, then the solution becomes

\begin{equation}\label{T-A1}
T(\tau)=T_0\Bigl(\frac{\tau_0}{\tau}\Bigr)^{\frac{1}{3}}
\ep^{\frac{\gamma}{3}(\frac{1}{\tau_0}-\frac{1}{\tau})}
\,.
\end{equation}	
The exponential factor on the right hand side shows that the temperature decreases with the proper time more slowly in \eqref{T-A1} than 
in \eqref{Bjorken}. 

\section{Concluding Comments}

There is a large number of open problems associated with the circle of ideas explored in this paper. In this section we list a few.

(i) There is a conjecture that there may be a bound from below on the ratio of the bulk viscosity to the shear viscosity \cite{sashab}

\begin{equation}\label{sashab}
 2\Bigl(\frac{1}{3}-C_s^2\Bigr)\leq \frac{\zeta}{\eta}
\,.
\end{equation}
Here $C_s$ is the speed of sound. For the fuzzy bag it is simply $C_s^2=(2T^2-\alpha)/(6T^2-\alpha)$ \cite{ads}. Combining this 
with \eqref{svis} and \eqref{bvis}, we learn 

\begin{equation}\label{bound2}
	\frac{T^2-k\alpha}{6T^2-\alpha}\leq\frac{3m}{4\kappa}
\,.
\end{equation}

Thus, if we accept the idea that the $\alpha$-corrections are small, we can immediately deduce from \eqref{kss} and \eqref{bound2} that for $m\approx 0.024$ the (temperature-independent) ratio $\eta/s$ is limited to 

\begin{equation}\label{window}
0.08\lesssim\eta/s\lesssim 0.11
\,.
\end{equation}
This is a very narrow interval even in the temperature range of interest, where \eqref{eofs} is valid. Is it reasonable? If not, what will happen with the bounds?

(ii) Given the solution \eqref{T}, we can describe how it behaves for large proper times.\footnote{Clearly, this is of academic interest only.} To leading order, we have

\begin{equation}\label{Tlong}
T(\tau)=\sqrt{\frac{\alpha}{2}}+\frac{1}{\alpha}\frac{c}{\tau}+O(\tau^{-2})
\,.
\end{equation}
This result shows that a small $\alpha$ expansion is no more appropriate. 

Interestingly, the temperature approaches the value $\sqrt{\alpha/2}$ which is, according to \cite{az2}, 
the critical temperature.\footnote{Note that for $\alpha=0.09\,\text{GeV}^2$ a simple estimate of $\sqrt{\alpha/2}$ 
yields $T_c=0.21\,\text{GeV}$ that looks quite satisfactory.} In other words, there is a non-zero limiting value of $T(\tau)$ which turns out 
to be $T_c$. It is worth mentioning that the critical slowing down was also observed in the MIT bag model \cite{kapusta-rev} under the assumption that the bulk viscosity has a power singularity at $T=T_c$ \cite{kerstin}. It seems that this is not the case for regular viscosity coefficients, as 
seen from the examples \eqref{conk}, \eqref{T-kapusta} and \eqref{T-A1}.

The above conclusion requires a caveat. The fuzzy bag pressure \eqref{eofs} is negative at $T_c$. The 
reason for this is very simple and is contained in the formula \eqref{eofs}. It is valid for $T\geq T_{\text{\tiny min}}$, 
where $T_{\text{\tiny min}}$ is close but not exactly equal to $T_c$. So, this doesn't mean that cavitation occurs. It means that the model for the pressure (together with those for the viscosities) needs to be refined. It would be interesting to see what happens if one does so.\footnote{A step in this direction was taken recently in \cite{sree}, where the lattice result of \cite{baz} was used at low ($T\lesssim 0.3\,\text{GeV}$) temperatures.}

(iii) While the lattice data for pure gauge theories suggest that at high temperatures the equation of state of the original bag model does get modified  
by the $T^2$ term \cite{lattice-K,pan}, the situation with the equation of state in $2+1$ flavor QCD is far from perfect. There is a discrepancy between the data recently reported by the Wuppertal-Budapest collaboration \cite{fodor} and those of the HOT-QCD collaboration \cite{baz}. While this issue is being or will be resolved by further lattice simulations, it is tempting to consider a general polynomial model for the pressure

\begin{equation}\label{zf}
	p(T)=a(T^4+8\gamma T^3-\alpha T^2+4\beta T)-B
	\,.
\end{equation}
This is a $3$-parameter deformation of the original bag model which reduces to the fuzzy bag \eqref{eofs} for $\beta=\gamma=0$.

In the absence of dissipative effects, we can obtain an explicit formula for $T(\tau)$ by solving a cubic equation. For small $\tau$ it is given by 

\begin{equation}\label{zfT}
T(\tau)=-2\gamma+f^{\frac{1}{3}}
\biggl[\Bigl(\oh+\oh\sqrt{1-\frac{(\alpha+24\gamma^2)^3}{54 f^2}}\Bigr)^{\frac{1}{3}} 
+\Bigl(\oh-\oh\sqrt{1-\frac{(\alpha+24\gamma^2)^3}{54 f^2}}\Bigr)^{\frac{1}{3}}\biggr]
\quad ,
\end{equation}
where $f=c\tau^{-1}-16\gamma^3-\alpha\gamma-\beta$ and $c=\tau_0(T^3_0+6\gamma T_0^2-\oh\alpha T_0+\beta)$. The solution \eqref{T} emerges in the limit $\beta ,\gamma\rightarrow 0$.

In Figure 5 we give an example of the proper time evolution of the temperature as it follows from \eqref{zfT}. We fit the lattice data for the interaction measure \cite{fodor} to the polynomial model \eqref{zf}, as shown in the left panel. This yields $a=2.8$, $\gamma=6.5\,\text{MeV}$, $\alpha=0.09\,\text{GeV}^2$, $\beta=(0.14\,\text{GeV})^3$, and $B=(0.2\,\text{GeV})^4$.\footnote{In doing so, we take the same values of $\alpha$ and $B$ as in the fuzzy bag model.} As before, we take $T_0=0.8\,\text{GeV}$ and $\tau_0=0.08\,\text{fm/c}$. What we see from the right panel is that 
the temperature decreases with the proper time more slowly in the presence of the additional terms in the equation of states. Moreover, a difference between the deformations of the equation of state becomes visible on the temperature curves when $\tau\gtrsim 1\,\text{fm/c}$.

\begin{figure}[ht]
\centering
\includegraphics[width=7.1cm]{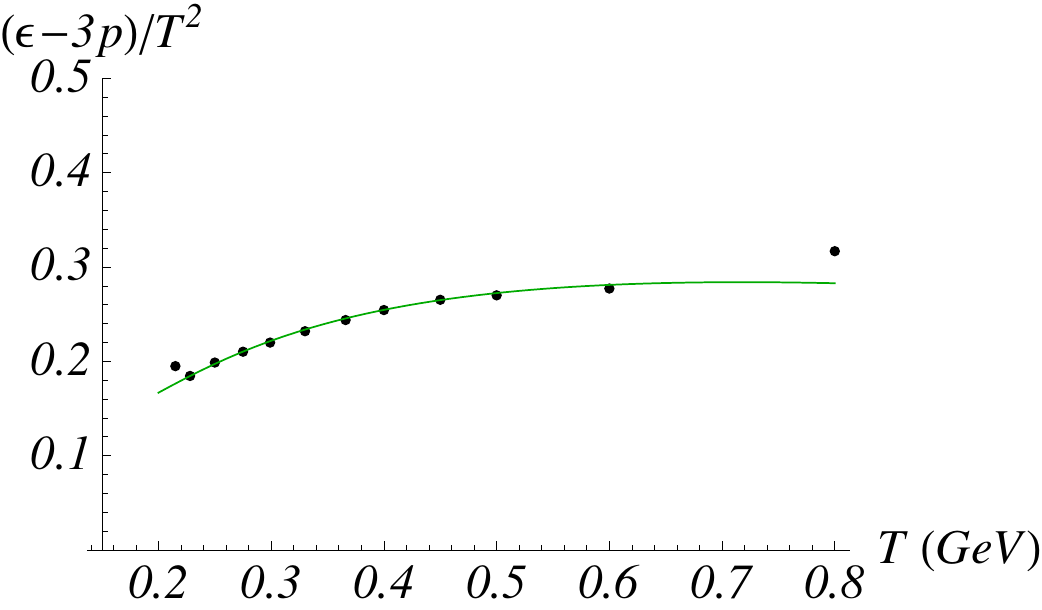}
\hfill
\includegraphics[width=6.9cm]{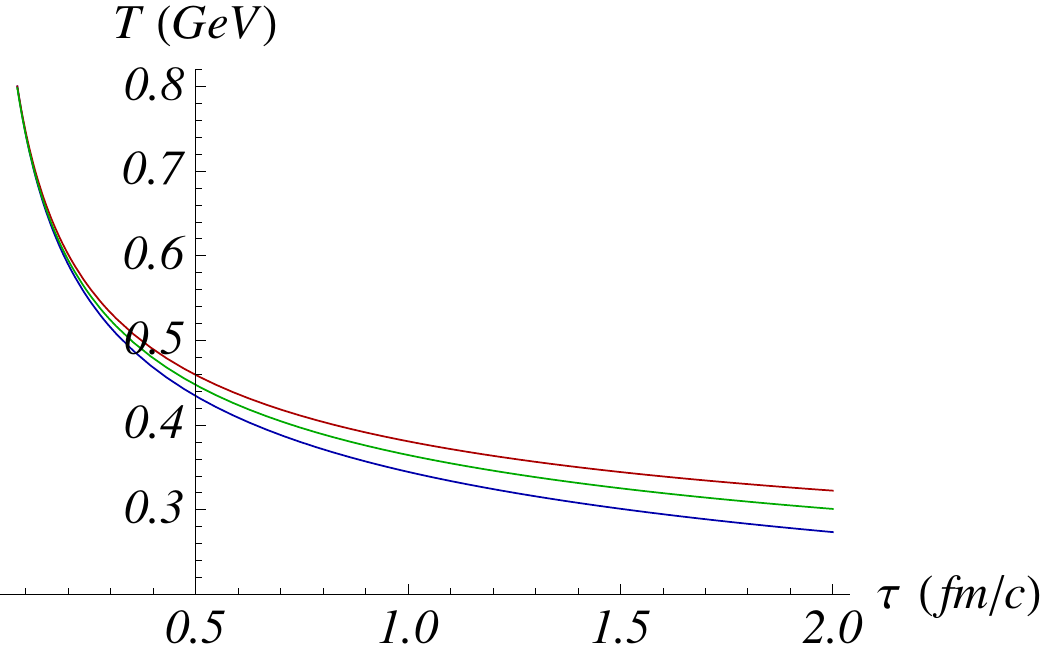}
\caption{\small{Left: The interaction measure normalized by $T^2$. The curve corresponds to \eqref{zf}. The dots come from the data \cite{fodor} at $R=28.15$ (physical quark masses). Right: Evolution of the temperature. As in Figure 1, the lower blue and upper red curves correspond to the MIT and fuzzy bag models. The middle curve represents \eqref{zfT}.}} 
\end{figure}

(iv) Here we used the simplified model for a boost-invariant expanding fluid which allowed us to find the solutions to the hydrodynamic equations in the first order theory. The main reasons for doing so were to find the solutions analytically and to avoid the issue of uncertainty about the transport (relaxation) coefficients at higher orders. Although the model is a good approximation to the early time evolution of the QCD matter, it would be very interesting to see whether our findings can stand the test of a $3+1$-dimensional hydrodynamic code.

\begin{acknowledgments}
We thank S. Hofmann for suggesting that we look into the hydrodynamics equations for a boost-invariant expanding 
fluid with the equation of state taken from the fuzzy bag model and for stimulating discussions. We also wish to thank J. Kapusta 
for reading the manuscript and Z. Fodor for drawing our attention to his work. This work was supported in part by DFG "Excellence Cluster" and the Alexander von Humboldt Foundation.
\end{acknowledgments}

\appendix
\begin{center}
{\bf APPENDIX}
\end{center}
\renewcommand{\theequation}{A.\arabic{equation}}
\setcounter{equation}{0}
The purpose of this appendix is to describe the evolution of the energy density, pressure and trace anomaly in the absence of 
dissipative effects. The corresponding calculations are trivial but a little tedious.

Combining the solution \eqref{T} with the equation of state \eqref{eofs}, we obtain

\begin{widetext}

	\begin{equation}\label{E}
		\varepsilon(\tau)=
		\begin{cases}
		3a\biggl[\bigl(\frac{c}{\tau}\bigr)^{\frac{2}{3}}
		\Bigl[\Bigl(\oh+\oh\sqrt{1-\frac{\alpha^3}{54}\bigl(\frac{\tau}{c}\bigr)^2}\Bigr)^{\frac{1}{3}} 
		+\Bigl(\oh-\oh\sqrt{1-\frac{\alpha^3}{54}\bigl(\frac{\tau}{c}\bigr)^2}\Bigr)^{\frac{1}{3}}\Bigr]^{-1}
		+\frac{1}{3}\alpha\biggr]^2-\frac{1}{12}a\alpha^2+B	
		& \text{if $\tau\leq\sqrt{\frac{54}{\alpha^3}}c$}
		\,,\\
		3a\biggl[\sqrt{\frac{3}{2\alpha}}\bigl(\frac{c}{\tau}\bigr)
		\cos^{-1}\Bigl[\frac{1}{3}\arccos\Bigl(\sqrt{\frac{54}{\alpha^3}}\frac{c}{\tau}\Bigr)\Bigr]+\frac{1}{3}\alpha
		\biggr]^2-\frac{1}{12}a\alpha^2+B
		&\text{if $\tau\geq\sqrt{\frac{54}{\alpha^3}}c$ }\,	
\end{cases}
		\end{equation}
and 
\begin{equation}\label{p}
	p(\tau)=
	\begin{cases}
		a\bigl(\frac{c}{\tau}\bigr)^{\frac{4}{3}}
		\Bigl[\Bigl(\oh+\oh\sqrt{1-\frac{\alpha^3}{54}\bigl(\frac{\tau}{c}\bigr)^2}\Bigr)^{\frac{1}{3}} 
		+\Bigl(\oh-\oh\sqrt{1-\frac{\alpha^3}{54}\bigl(\frac{\tau}{c}\bigr)^2}\Bigr)^{\frac{1}{3}}\Bigr]^{-2}
		-\frac{1}{4}a\alpha^2-B	& \text{if $\tau\leq\sqrt{\frac{54}{\alpha^3}}c$}\,,\\
	\frac{3a}{2\alpha}\bigl(\frac{c}{\tau}\bigr)^2
	\cos^{-2}\Bigl[\frac{1}{3}\arccos\Bigl(\sqrt{\frac{54}{\alpha^3}}\frac{c}{\tau}\Bigr)\Bigr]
	-\frac{1}{4}a\alpha^2-B	
&\text{if $\tau\geq\sqrt{\frac{54}{\alpha^3}}c$ }\,.
	\end{cases}
	\end{equation}

\end{widetext}

It is instructive to look at the expansion of $\varepsilon (\tau)$ in powers of $\tau$. To first order, it is given by  

\begin{equation}\label{Eshort}
			\varepsilon(\tau)=3a\Bigl(\frac{c}{\tau}\Bigr)^{\frac{4}{3}}
			+a\alpha\Bigl(\frac{c}{\tau}\Bigr)^{\frac{2}{3}}+O(1)
			\,.
		\end{equation}
In this formula the second term is the leading $\alpha$-correction to the evolution of the energy density of the MIT bag model \cite{bj}

\begin{equation}\label{Ashort}
			\varepsilon(\tau)=3a\Bigl(\frac{c}{\tau}\Bigr)^{\frac{4}{3}}\,.
		\end{equation}
Because it is positive, the energy density of the fuzzy bag falls faster. This explains the curves in the right panel of Figure 1. Moreover, we can get an idea about the evolution of the trace anomaly (interaction measure) in this time interval. It is 

\begin{equation}\label{anomaly}
\frac{\varepsilon-3p}{T^4}(\tau)=2a\alpha\Bigl(\frac{\tau}{c}\Bigr)^{\frac{2}{3}}+O(1)
\,.
\end{equation}

Actually, the above analysis can be easily extended to the solution \eqref{T-A} by replacing $c$ with $\hat c\,\ep^{-\frac{\gamma}{\tau}}$.



\begin{thebibliography}{99}
\bibitem{dau}
L.D. Landau, Izv. Akad. Nauk Ser. Fiz. {\bf 17}, 51 (1953).
\bibitem{rev}
H. St\"ocker and W. Greiner, Phys.Rep. {\bf 137}, 227 (1986); U.W. Heinz, arXiv:0901.4355 [nucl-th]; M. Luzum, arXiv:0908.4100 [nucl-th]; 
P. Romatschke, Int.J.Mod.Phys. {\bf E19}, 1 (2010).
\bibitem{wid}
R. Baier, P. Romatschke and U. Wiedemann, Phys.Rev.C {\bf 73}, 064903 (2006).
\bibitem{pis}
R.D. Pisarski, Prog.Theor.Phys.Suppl. {\bf 168}, 276 (2007).
\bibitem{lattice-K}
G. Boyd {\it et al.}, Phys.Rev.Lett. {\bf 75}, 4169 (1995);
\bibitem{T2}
P.N. Meisinger, T.R. Miller, and M.C. Ogilvie, Phys.Rev.D {\bf 65}, 034009 (2002).
\bibitem{lattice}
 M. Cheng {\it et al.}, Phys.Rev.D {\bf 77}, 014511 (2008).
\bibitem{pan} 
M. Panero, Phys.Rev.Lett. {\bf 103}, 232001 (2009).
\bibitem{baz}
 A. Bazavov {\it et al.}, Phys. Rev.D {\bf 80}, 014504 (2009). 
\bibitem{q2-rev}
V.I. Zakharov, arXiv:1010.4482 [hep-ph].
\bibitem{ads}
O. Andreev, Phys.Rev.D {\bf 76}, 087702 (2007).
\bibitem{bj}
J.D. Bjorken, Phys.Rev.D {\bf 27}, 140 (1983).
\bibitem{kapustaMc}
J. Kapusta, L. McLerran, and D.K. Srivastava, Phys. Lett.B {\bf 283}, 145 (1992).
\bibitem{A-q2}
O. Andreev, Phys.Rev.D {\bf 73}, 107901 (2006); O. Andreev and V.I. Zakharov, Phys.Rev.D {\bf 74}, 025023 (2006).
\bibitem{kss}
P.K. Kovtun, D.T. Son, and A.O. Starinets, Phys.Rev.Lett. {\bf 94}, 111601 (2005).
\bibitem{zajc}
W.A. Zajc, Nucl.Phys. {\bf A805}, 283 (2008).
\bibitem{roma}
P. Romatschke and U. Romatschke, Phys.Rev.Lett. {\bf 99}, 172301 (2007).
\bibitem{Bv}
H. Kouno, M. Maruyama, F. Takagi, and K. Saito, Phys.Rev.D {\bf 41}, 2903 (1990); A. Muronga, Phys.Rev.Lett. {\bf 88}, 062302 (2002).
\bibitem{kapusta-rev}
J.I. Kapusta, arXiv:0809.3746 [nucl-th].
\bibitem{lattice-shear}
A. Nakamura and S. Sakai, Phys.Rev.Lett. {\bf 94}, 072305 (2005); H.B. Meyer, Phys.Rev.D {\bf 76}, 101701 (2007).
\bibitem{kapusta-shear}
J.I. Kapusta and T. Springer, Phys.Rev.D {\bf 78}, 066017 (2008).
\bibitem{DimaK}
D. Kharzeev and K. Tuchin, J.High Energy Phys. 0809 (2008) 093.
\bibitem{lattice-bulk}
H.B. Meyer, Phys.Rev.Lett. {\bf 100}, 162001 (2008).
\bibitem{raj}
K. Rajagopal and N. Tripuraneli, J.High Energy Phys. 018 (2010) 1003. 
\bibitem{sashab}
A. Buchel, Phys.Lett.B {\bf 663}, 286 (2008).
\bibitem{az2}
O. Andreev and V.I. Zakharov, Phys.Lett.B {\bf 645}, 437 (2007).
\bibitem{kerstin}
K. Paech and S. Pratt, Phys.Rev.C {\bf 74}, 014901 (2006).
\bibitem{sree}
J.R. Bhatt, H. Mishra, and V. Sreekanth, J.High Energy Phys. 1011 (2010) 106.
\bibitem{fodor}
S. Bors\'anyi {\it et al.}, J.High Energy Phys 1011 (2010) 077.

\end{thebibliography}
\end{document}